\documentclass[prl,twocolumn,amsmath,amssymb,superscriptaddress,longbibliography]{revtex4-1}

\usepackage{graphicx} 
\graphicspath{{figures/}}

\usepackage{dcolumn} 
\usepackage{color}
\usepackage{amsmath} 
\usepackage{braket}
\usepackage[breaklinks,colorlinks,bookmarks=false,citecolor=blue,linkcolor=red,urlcolor=blue]{hyperref}
\usepackage{cleveref} 
\usepackage{times} 
\crefname{equation}{Eq.}{Eqs.}
\Crefname{equation}{Eq.}{Eqs.}  
\crefname{figure}{Fig.}{Figs.}
\Crefname{figure}{Fig.}{Figs.}  
\crefname{section}{section}{sections}
\Crefname{section}{Section}{Sections}

\begin{document}

\title{Valence bond solid and possible deconfined quantum criticality in an 
extended kagome lattice Heisenberg antiferromagnet}

\author{Alexander Wietek} 
\email{awietek@flatironinstitute.org}
\affiliation{Center for Computational Quantum Physics, Flatiron
  Institute, 162 5th Avenue, NY 10010, New York, USA}

\author{Andreas M. L\"auchli} 
\affiliation{Institut f\"ur Theoretische Physik, Universit\"at Innsbruck, A-6020 Innsbruck, Austria}

\date{\today}

\begin{abstract}
  We present numerical evidence for the existence of an extended
  valence bond solid (VBS) phase at $T=0$ in the kagome 
  $S=1/2$ Heisenberg antiferromagnet with ferromagnetic further-neighbor
  interactions. The VBS is located at the boundary between two
  magnetically ordered regions and extends close to the
  nearest-neighbor Heisenberg point. It exhibits a diamond-like
  singlet covering pattern with a $12$-site unit-cell. Our results suggest the
  possibility of a direct, possibly continuous, quantum phase transition from
  the neighboring $\mathbf{q}=0$ coplanar magnetically ordered phase into
  the VBS phase. Moreover, a second phase which breaks lattice symmetries,
  and is of likely spin-nematic type, is found close to the transition to the
  ferromagnetic phase. The results have been obtained using large-scale 
  numerical Exact Diagonalization. We discuss implications of our results on the
  nature of nearest-neighbor Heisenberg antiferromagnet.
\end{abstract}

\maketitle

\paragraph{Introduction ---}
We expect the unexpected when strong electron interactions meet
geometric frustration. The emergence of novel exotic states of matter
in frustrated quantum magnets is intensely studied in experiments,
theory, and numerical computations. Several materials and theoretical
models exhibit a lack of magnetic ordering even at lowest
temperatures. Instead, genuine quantum many-body states, like quantum
spin liquids~\cite{balents2010,Savary2016} or valence bond solids
(VBS) can be
observed~\cite{Majumdar1969,Fouet2003,Mambrini2006,Zhitomirsky1996,Iqbal2011b}.
Several experiments also have given evidence for emerging VBS
phases in a variety of
materials~\cite{Matan2010,Ruegg2010,Sheckelton2012,Smaha2019}.


The nearest-neighbor kagome lattice Heisenberg spin $1/2$
antiferromagnet arguably remains one of the most puzzling conundrum
in frustrated magnetism.  Various scenarios on the nature of its
ground state have been proposed.  It has been found early, that a VBS
is energetically
competitive~\cite{Marston1991,Leung1993,Nikolic2003,Singh2007,Capponi2013}.
However, more recent numerical studies suggest, that different spin
disordered states are a more likely scenario. Several density-matrix
renormalization group (DMRG) studies later suggested the possibility
of a gapped spin liquid ground
state~\cite{Yan2011,Depenbrock2012}. More recently, variational Monte
Carlo and tensor network studies also suggested a gapless spin liquid
state being
realized~\cite{Ran2007,Iqbal2015,Iqbal2011,Iqbal2013,He2017,Liao2017}.
While conclusion on the nature of its ground state has not unanimously
been reached to date~\cite{Laeuchli2019}, several exotic new states of
matter have been clearly identified in close proximity to the
nearest-neighbor model~\cite{Bauer2014,Gong2014, Wietek2015,He2014}.
Among those, a chiral spin liquid has been found in an extended
Heisenberg model with antiferromagnetic second and third
nearest-neighbor interactions~\cite{Gong2014,Wietek2015,He2014}.  The
classical ground state phase diagram of this model has previously been
established~\cite{Messio2011,Messio2012}. A phase transition between
two magnetic orders has been found for antiferromagnetic
interactions. In the quantum case, the chiral spin liquid phase is
located along the transition line between these two magnetic phases and
extends close to the nearest-neighbor point.  The classical phase
diagram also contains a phase transition line between two types of
coplanar magnetic orders for {\em ferromagnetic} second and third
nearest-neighbor interactions.  Given that some frustrated kagome
materials involving both ferromagnetic and antiferromagnetic couplings
are known to exist~\cite{Fak2012,Kermarrec2014,Iqbal2016}, there is a
strong interest to explore whether novel phases also emerge at or in
the vicinity of the classical transition line at $J_3=2 J_2<0$.

\begin{figure}[b]
  \centering \includegraphics[width=\linewidth]{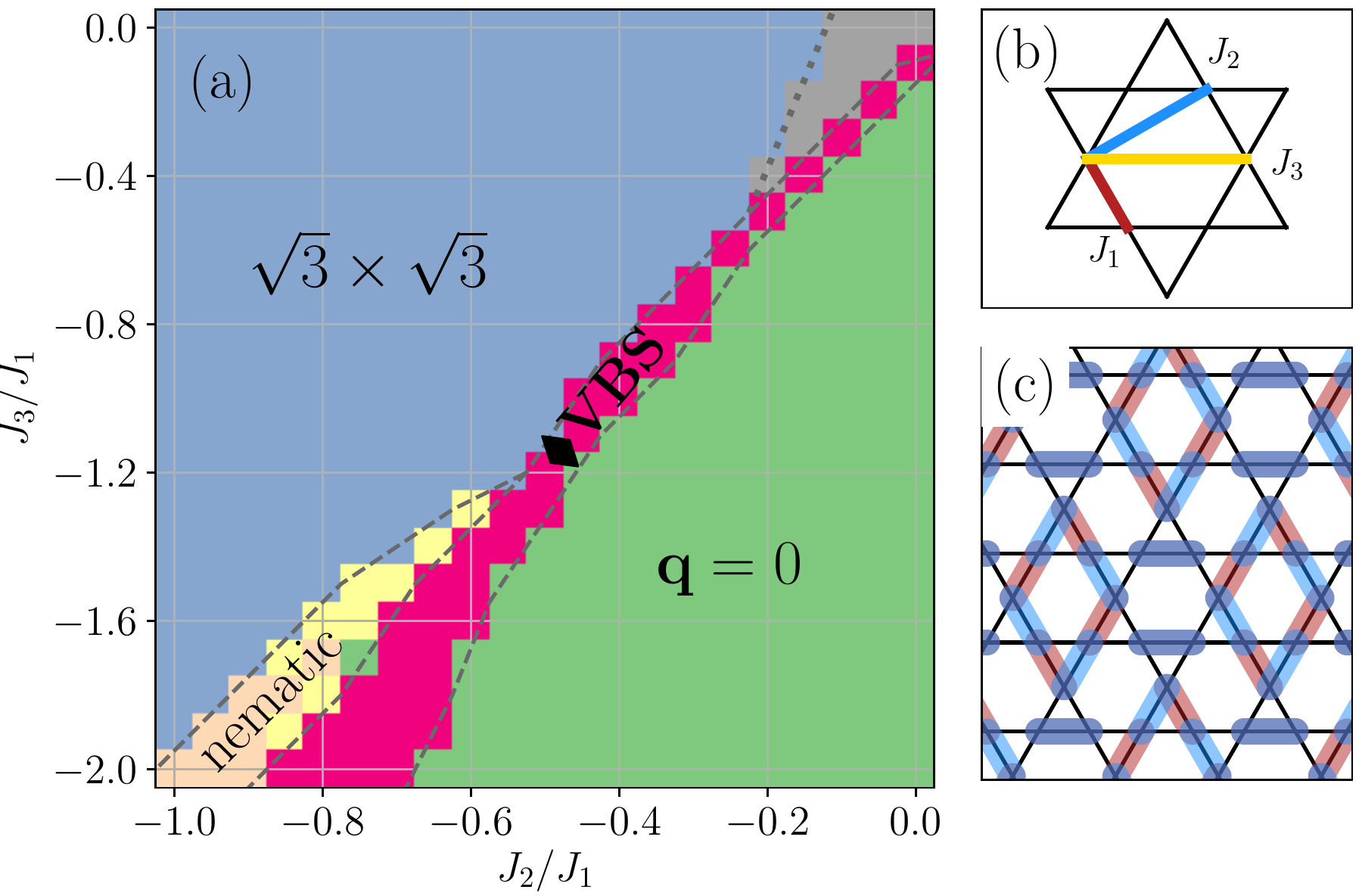}
  \caption{(a) Approximate phase diagram of the extended kagome
    Heisenberg model \cref{eq:hamiltonian} for $J_1>0$ and
    $J_2, J_3 \leq 0$ as obtained from ED on a 36-site simulation
    cluster. Between two regions of magnetic $\mathbf{q}=0$ and
    $\sqrt{3} \times \sqrt{3}$ order a diamond VBS and a spin nematic
    phase are emerging. Different colors denote the quantum numbers of
    the first excited state.  Green: $S=1$, $\Gamma$.D6.A2 or
    $\Gamma$.D6.E2. Blue: $S=1$, $\Gamma$.D6.B1 or K.D3.A1.  Pink:
    $S=0$, M.D2.A2. Orange: $S=0$, M.D2.A1. Yellow: $S=2$,
    $\Gamma$.D6.A1. Gray: $S=0$, various space group sectors. Gray
    lines are a guide to the eye. (b) Coupling geometry for the
    Hamiltonian \cref{eq:hamiltonian}. (c) Structure of the diamond
    VBS with a 12-site unit cell. Dimer coverings on the diamond
    structure are in resonance.}
  \label{fig:phasediagram}
\end{figure}


Here, we investigate the kagome spin $1/2$ Heisenberg antiferromagnet
with additional ferromagnetic second and third nearest-neighbor
interactions. We present conclusive numerical evidence for the
appearance of a {\em diamond} VBS phase in an extended parameter
range.  The VBS phase is located in the vicinity of the classical
transition line between the $\mathbf{q}=0$ and
$\sqrt{3} \times \sqrt{3}$ magnetic orders.  Interestingly, the phase
extends close up to the nearest-neighbor Heisenberg point.

\begin{figure*}[t!]
  \centering \includegraphics[height=0.54\columnwidth]{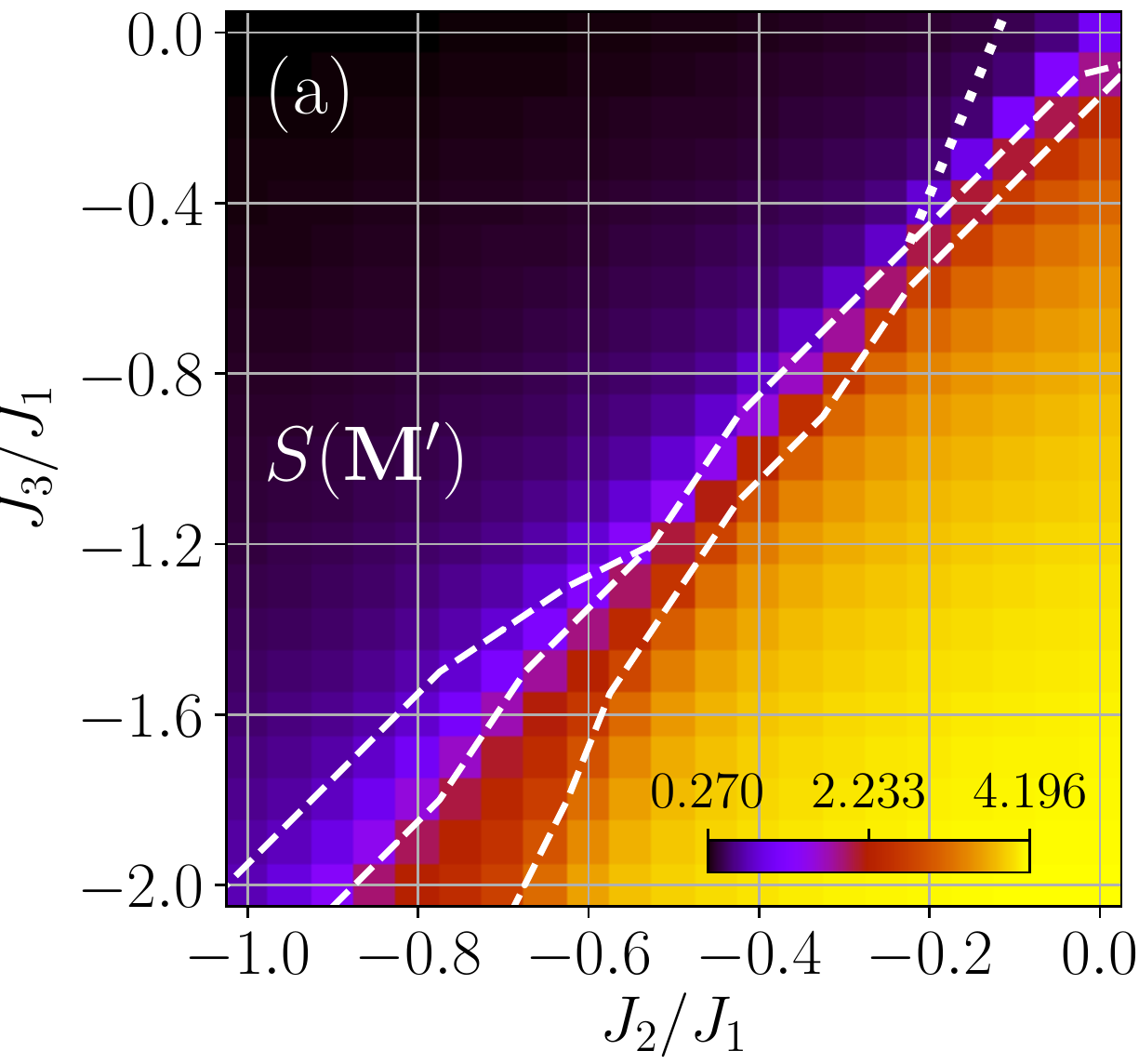}
  \includegraphics[height=0.54\columnwidth]{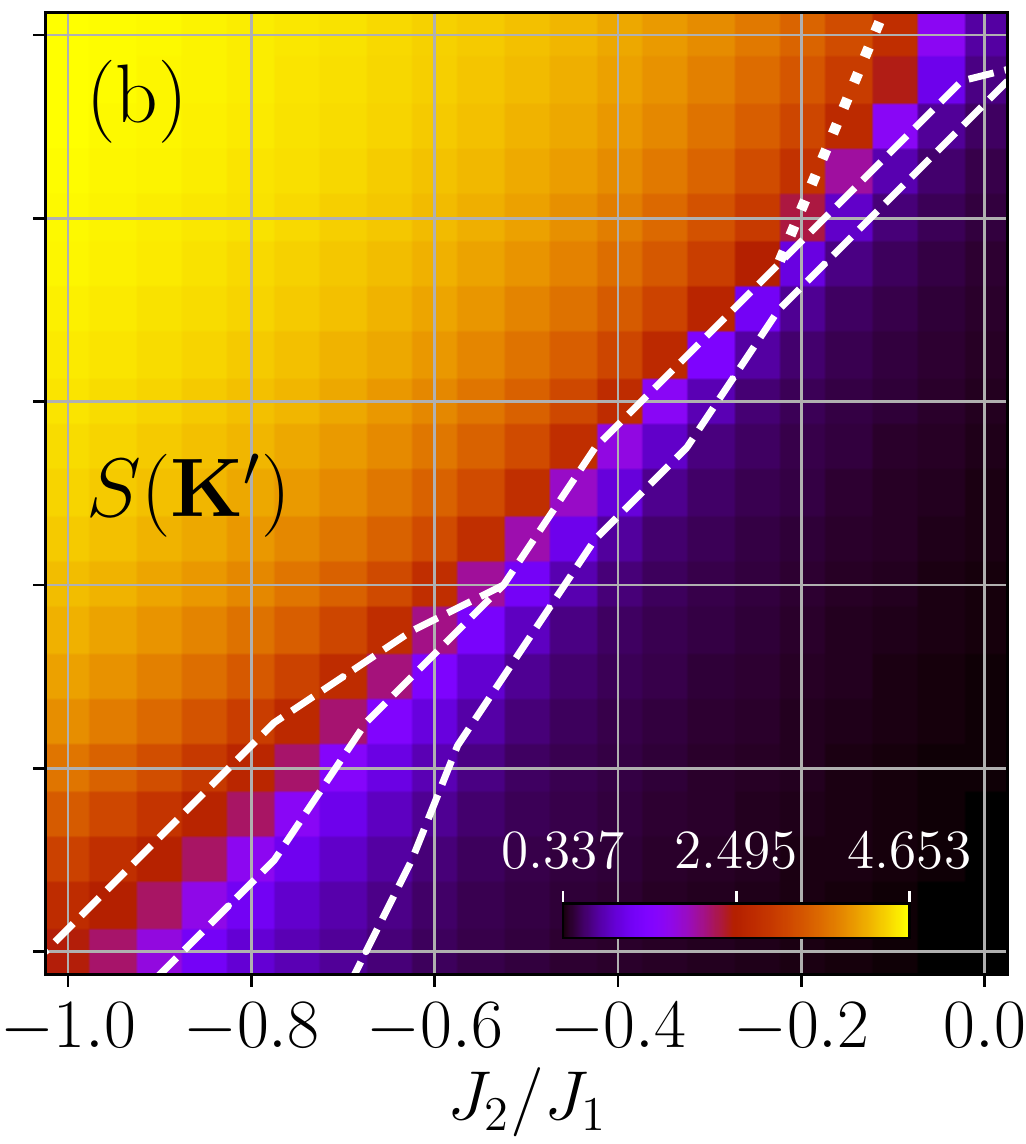}
  \includegraphics[height=0.54\columnwidth]{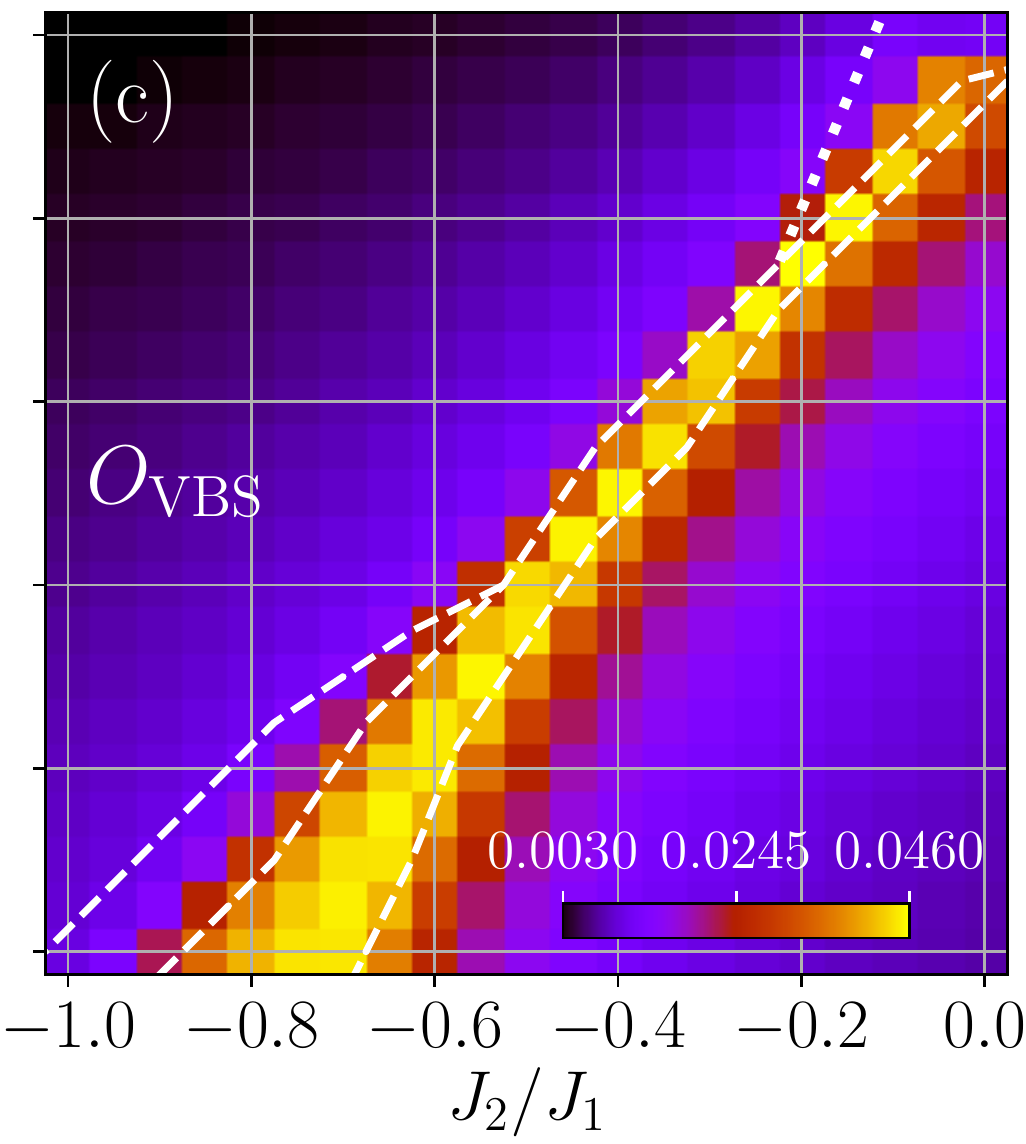}
  \includegraphics[height=0.54\columnwidth]{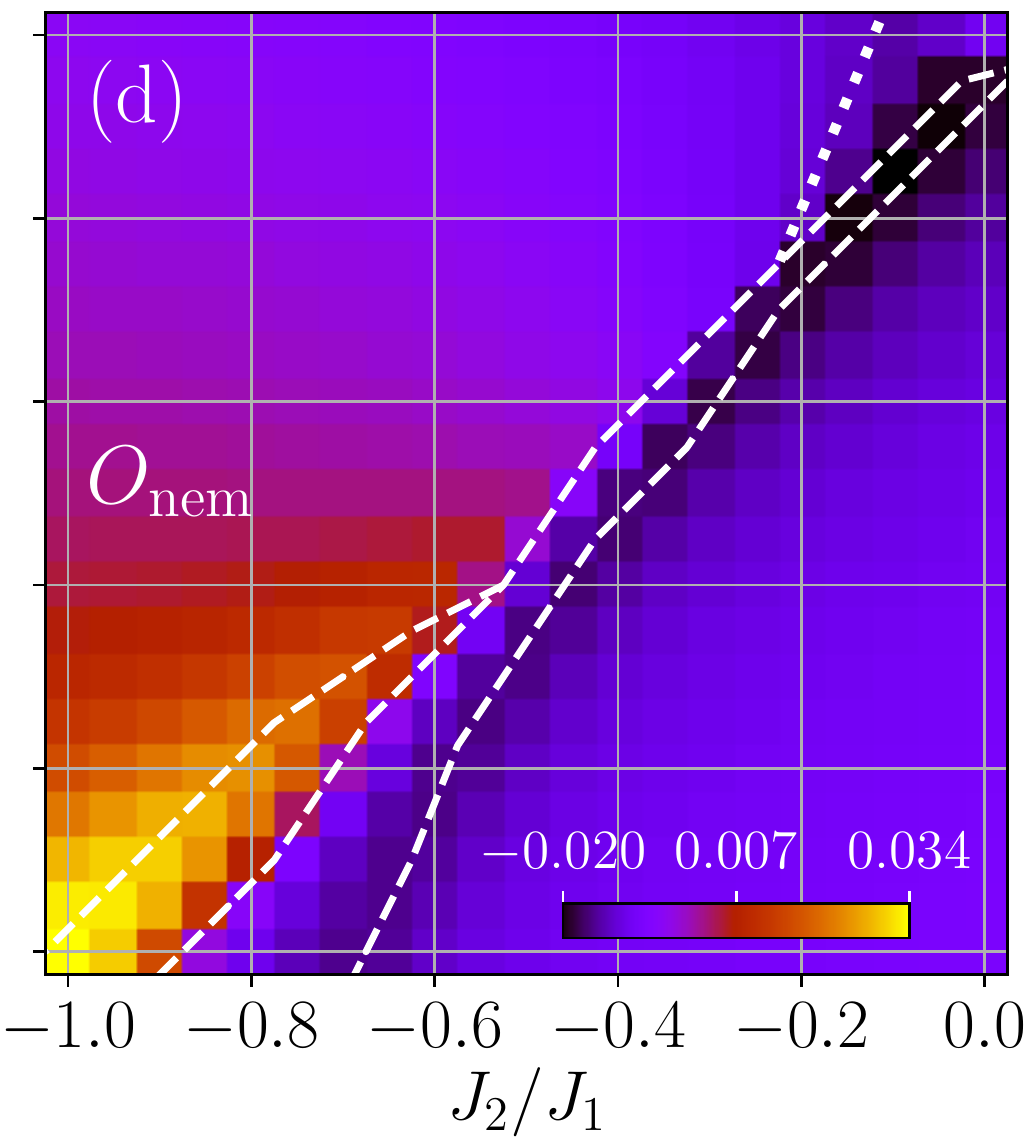}
  \caption{ED results on a 36-site simulation
    cluster: (a) Spin structure factor $S(\textrm{\textbf{M}}^\prime)$
    evaluated at $\textrm{\textbf{M}}^\prime= (2\pi, 2\pi/\sqrt{3})$,
    indicating $\mathbf{q}=0$ magnetic order.  (b) Spin structure
    factor $S(\textrm{\textbf{K}}^\prime)$ evaluated at
    $\textrm{\textbf{K}}^\prime = (8 \pi / 3, 0)$, indicating
    $\sqrt{3} \times \sqrt{3}$ magnetic order. (c) Diamond VBS order
    parameter $O_{\text{VBS}}$ as defined in
    \cref{eq:vbsorderparameter}. A diamond VBS phase is emerging in
    between the two magnetic orders. (d) Nematic phase order parameter
    $O_{\text{nem}}$ as defined in \cref{eq:nematicorderparameter},
    indicating the extent of the plaquette-nematic phase. Note the
    good agreement between the order parameter inferred phase diagram
    and the excited state spectroscopy phase diagram of
    \cref{fig:phasediagram} indicated with the dashed white lines.}
  \label{fig:orderparameters}
\end{figure*}

\paragraph{Model and phase diagram ---}
We consider the Hamiltonian,
\begin{align}
  \label{eq:hamiltonian}
  \begin{split}
    H = &\ J_1\sum\limits_{\langle i, j \rangle} \vec{S}_i \cdot
    \vec{S}_j +
    J_2\sum\limits_{\langle\langle i, j \rangle\rangle}  \vec{S}_i \cdot \vec{S}_j  \\
    &+ J_3\sum\limits_{\langle\langle\langle i, j
      \rangle\rangle\rangle_{\textrm{h}}} \vec{S}_i \cdot \vec{S}_j,
  \end{split}
\end{align}
on a kagome lattice geometry, where
$\vec{S}_i = (S^x_i, S^y_i, S^z_i)$ denotes spin $1/2$ operators,
$\langle \ldots \rangle$ and $\langle\langle \ldots \rangle\rangle$
denotes the sum over nearest- and second nearest-neighbor sites, and
$\langle\langle\langle \dots \rangle\rangle\rangle_{\textrm{h}}$
denotes sum over third nearest-neighbor interactions only across the
hexagons of the kagome lattice, cf. \cref{fig:phasediagram}(b). In the
following, we set $J_1=1$ and focus on the case of ferromagnetic
couplings $J_2 < 0$ and $J_3 < 0$.

Most of our results are obtained by Exact Diagonalization (ED) calculations on
a $N=36$ site kagome lattice with periodic boundary
conditions~\cite{Wietek2018,Laeuchli2011}. Its Brillouin zone features the
\textbf{K} and \textbf{M} points and is hence suited to stabilize both
the $\sqrt{3} \times \sqrt{3}$ and $\mathbf{q}=0$ order.  Selected results have
been obtained on smaller clusters as well as on the larger $N=48$ cluster~\cite{Laeuchli2019,Wietek2018}.
We detect ordering by investigating suitably chosen order parameters and
performing tower-of-states analysis, i.e. comparing quantum numbers of
finite-size energy eigenstates with theoretical predictions. The order
parameters of the ground state and finite-size energy spectra are
calculated on a grid for $J_2 \in [-1,0]$ with spacing $0.05$ and
$J_3 \in [-2,0]$ with spacing $0.1$.

For classical Heisenberg spins the phase diagram of this model has
been established in Ref.~\cite{Messio2012}. The
$\sqrt{3} \times \sqrt{3}$ magnetic phase is separated from the
$\mathbf{q}=0$ magnetic phase by a transition line located at $J_3 =
2J_2$. For $J_3<-2$ and $J_2<-1$ a ferromagnetic state is stabilized.


In \cref{fig:phasediagram} we present a first exploration of the
quantum ($S=1/2$) phase diagram based on a map organized by the
quantum numbers of the first excitation above the ground state. The
assignment of the phases is performed based on a tower-of-states
analysis for different candidate phases. According to this rationale,
the blue region indicates the $\sqrt{3} \times \sqrt{3}$ magnetic
order, the green region indicates the $\mathbf{q}=0$ magnetic order and
the pink region the VBS phase. The nematic phase extends in the yellow
and orange region, where two different quantum numbers are the first
excitation.  The gray lines serve as a guide to the eye and determine
approximate phase boundaries. Apart from the expected
$\sqrt{3} \times \sqrt{3}$ and $\mathbf{q}=0$ coplanar magnetic order
phases, we find an unanticipated diamond VBS and a lattice symmetry
breaking spin nematic phase located in the vicinity of the classical
transition line. In \cref{fig:orderparameters} we corroborate the
spectroscopy picture with an analysis of corresponding order
parameters. The spin nematic phase extends close to the classical
ferromagnetic phase, while the VBS phase extends close to the
nearest-neighbor point. We now proceed to characterize the reported
phases in more detail.

\paragraph{Magnetic order ---}
The $\mathbf{q}=0$ and $\sqrt{3} \times \sqrt{3}$ magnetic phases
break spin rotational SU($2$) symmetry and exhibit patterns of
magnetic ordering shown in the supplementary
material~\cite{SuppMat}. We consider the static spin structure factor,
\begin{equation}
  \label{eq:structurefactor}
  S(\mathbf{q}) = \frac{1}{N}\sum_{k,l = 1}^N
  \text{e}^{-i\mathbf{q}\cdot(\mathbf{r}_k - \mathbf{r}_l)}
  \langle \vec{S}_k \cdot \vec{S}_l \rangle. 
\end{equation}
For the two magnetic orders, the structure factor is peaked at the
points
\begin{align}
  \textrm{\textbf{M}}^\prime =
  (2\pi, 2\pi/\sqrt{3}) \quad &\text{ for } \mathbf{q}=0 \text{ order,} \\
  \textrm{\textbf{K}}^\prime =
  (8 \pi / 3, 0)        \quad &\text{ for } \sqrt{3} \times \sqrt{3} \text{ order,}
\end{align}
in the extended Brillouin zone, cf.~\cite{Messio2011}.  Hence,
$S(\textrm{\textbf{M}}^\prime)$ and $S(\textrm{\textbf{K}}^\prime)$
shown in \cref{fig:orderparameters}(a) and (b) identify both magnetic
phases, respectively.  The regions where these structure factors are
peaked coincide with the blue and green regions in
\cref{fig:phasediagram}. The blue region in \cref{fig:phasediagram}(a)
is given by the points, where the first excitation is a triplet,
$S=1$, state with K.D3.A1 or $\Gamma$.D6.B1 space group quantum
numbers~\cite{SuppMat, Lecheminant1997}. In the green region in
\cref{fig:phasediagram}(a), the triplet states, $S=1$, with
$\Gamma$.D6.A2 and $\Gamma$.D6.E2 space group quantum numbers are the
first excitation. Thus, the spin structure factor and energy
spectroscopy yield consistent results on the extent of these two
phases.

\begin{figure}[t]
  \includegraphics[width=\columnwidth]{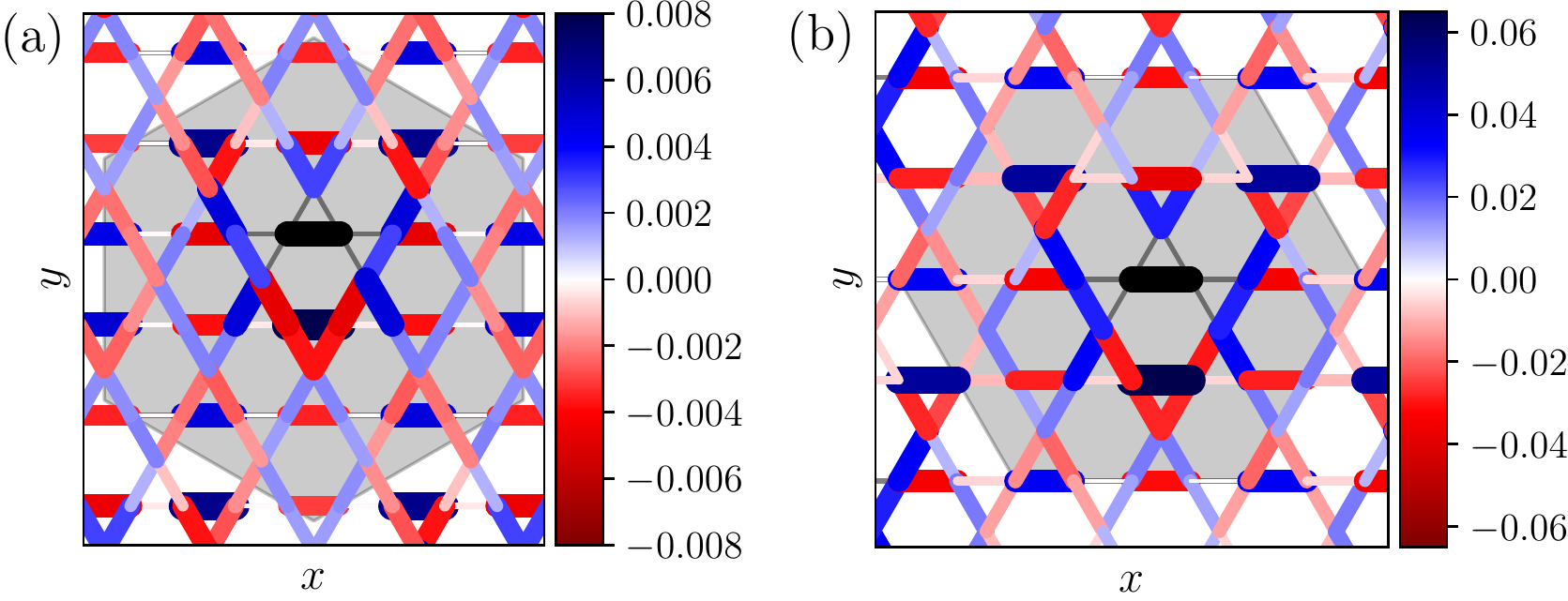}
  \caption{Ground-state dimer correlations in the VBS phase for
    $J_2/J_1 = -0.4$, $J_3/J_1 = -0.9$ from Exact Diagonalization (a):
    $S^z$ dimer-dimer correlations $D_{kl}^z$ as defined in
    Eq.~\ref{eq:szdimercorrs} on the $48$ site cluster (b):
    dimer-dimer correlations $D_{kl}$ as defined in
    Eq.~\ref{eq:dimercorrs} on the $36$ site cluster. The black line
    is used as the reference bond and the grey area shows the
    Wigner-Seitz simulation cell.}
  \label{fig:vbs_dimercorr_panel}
\end{figure}

\paragraph{Diamond VBS phase ---}
To identify the VBS and the lattice symmetry breaking spin-nematic
phase we consider the connected dimer correlations,
\begin{equation}
  \label{eq:dimercorrs}
  D_{kl} \equiv
  \langle ( \vec{S}_k \cdot \vec{S}_l) ( \vec{S}_1 \cdot \vec{S}_0)\rangle  -
  \langle \vec{S}_k \cdot \vec{S}_l\rangle
  \langle \vec{S}_1 \cdot \vec{S}_0\rangle,
\end{equation}
where the sites $0$ and $1$ are an arbitrary nearest-neighbor bond
chosen as reference.  These correlations are long-ranged in the VBS
phase and exhibit specific patterns of positive and negative
correlation that can be predicted for model VBS state.  For the $48$
site cluster we computed the diagonal $S^z$-dimer correlations
\begin{equation}
  \label{eq:szdimercorrs}
  D_{kl}^z \equiv
  \langle ( S^z_k \cdot S^z_l) ( S^z_1 \cdot S^z_0)\rangle  -
  \langle S^z_k \cdot S^z_l\rangle
  \langle S^z_1 \cdot S^z_0\rangle.
\end{equation}
The sign structure of these correlations serves as a first fingerprint
of the particular VBS phase realized. For the diamond VBS state the
expected sign structure of the dimer correlations are shown
in~\cite{SuppMat}. Thereby, we define an order parameter of the VBS
phase,
\begin{equation}
  \label{eq:vbsorderparameter}
  O_{\text{VBS}} \equiv \frac{1}{N}\sum_{\langle k,l \rangle} 
  \theta_{kl}^{\text{VBS}} D_{kl},  
\end{equation}
where $\theta_{kl}^{\text{VBS}} = \pm 1$ denotes the sign as defined
in the supplementary material.

This diamond VBS parameter $O_{\text{VBS}}$ is shown in
\cref{fig:orderparameters}(c), indicating the extent of the VBS
phase. It is located between the two magnetic orders and extends
basically along the whole classical transition line from $J_2=-1$ to
$J_2=0$. The region of pronounced $O_{\text{VBS}}$ also coincides with
the pink region in \cref{fig:phasediagram}. There, the first excited
state is a singlet $S=0$ state with M.D2.A2 space group quantum
number.

\begin{figure}[t]
  \includegraphics[width=\columnwidth]{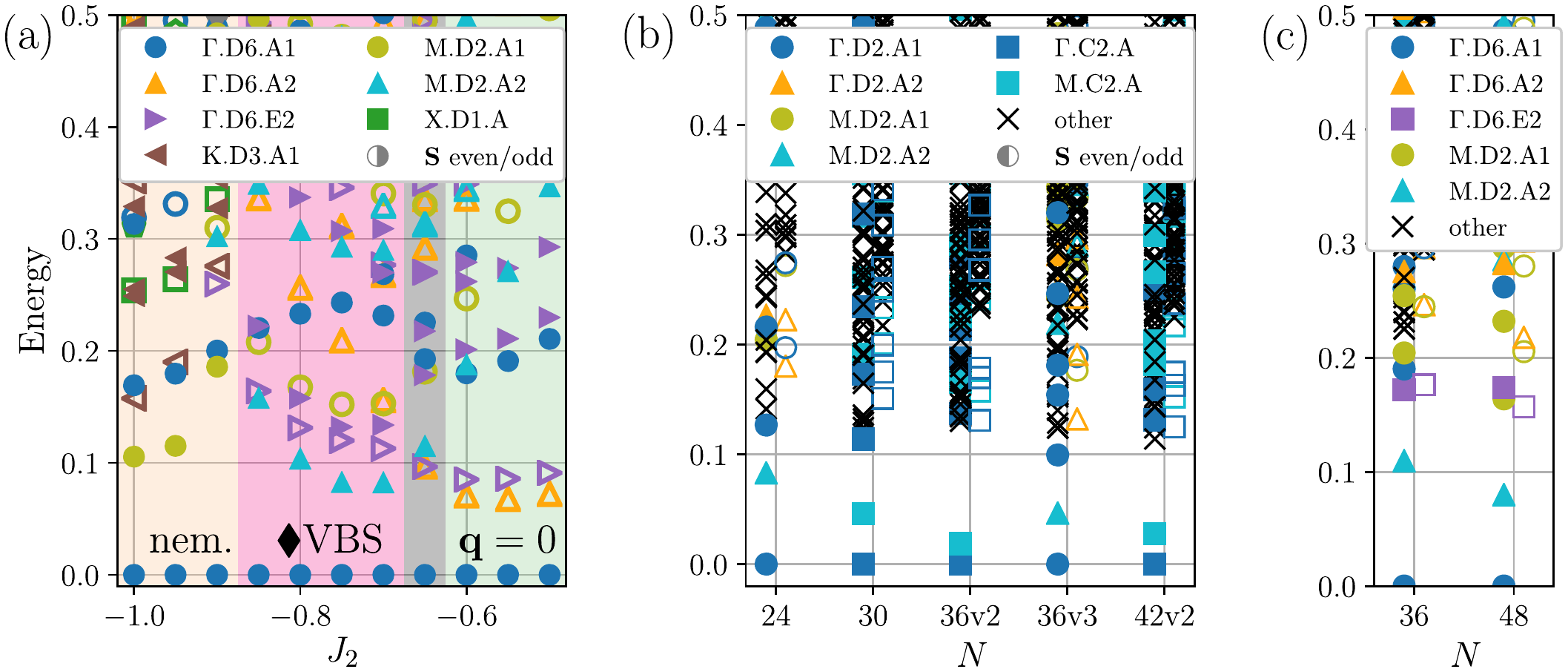}
  \caption{Diamond VBS: (a): energy spectra with quantum numbers for
    $J_3 = -2$ and $S_z=0$.  Different colors and symbols denote
    different quantum numbers. Full (empty) symbols denote even (odd)
    spin-flip symmetry eigenstates. The spin-nematic phase extends up
    to $J_2 \lesssim -0.9$ (orange shaded region). The first excited
    level is a singlet with space group quantum number M.D2.A1. The
    diamond VBS phase exists in the window (pink shaded region)
    $-0.9\lesssim J_2\lesssim -0.7$. The first excited state is a
    singlet with space group quantum number M.D2.A2. Beyond, the
    $\mathbf{q}=0$ is stabilized with triplet excitations (green
    shaded region). The narrow grey shaded region highlights the
    putative quantum critical N\'eel-VBS transition.  (b) energy
    spectra as a function of system size for $J_2=-0.16$ and
    $J_3=-0.4$.  The lowest excited level on all lattices is a singlet
    with momentum $M$. For a description of the used clusters, c.f.~Ref.~\cite{Laeuchli2011} 
    (c) Energy spectra for the $C_{6v}$ symmetric 36 and 48 site clusters at $J_2=-0.16$ and
    $J_3=-0.4$. The lowest excited states again have the same $M$ momentum and 
    space group sector.}
  \label{fig:vbs_spectra_panel}
\end{figure}

The precise nature of the reported VBS itself requires some more
care. There are two basic candidate VBS model states with a twelve
site unit cell~\cite{Yan2011,Poilblanc2011,Huh2011,Hwang2015}.  A
pinwheel VBS, where all dimers are static and the pinwheels all share
the same orientation. This particular state is eightfold degenerate, a
factor four from the translations and a factor two from the pinwheel
orientation. On the other hand, like in many other VBS scenarios,
there is a resonant version of this VBS, where we consider resonances
involving eight-site loops in the shape of a diamond lozenge. A fully
packed state of non-overlapping resonances is shown in
\cref{fig:phasediagram}(c). This state is actually twelve-fold
degenerate, a factor four from the translations, and a factor three
from the orientations of the diamond lozenges.  The dimer-dimer
correlations in these two model states are identical, so that dimer
correlations alone can not distinguish the two states. However the spectral
decomposition~\cite{SuppMat} reveals that beyond some common levels
the diamond VBS features a characteristic spin singlet $\Gamma$.D6.E2
level, while the pinwheel VBS comes with a characteristic
$\Gamma$.D6.A2 level. A close inspection of the energy spectrum of the
VBS phase in \cref{fig:vbs_spectra_panel}(a)\&(c) reveals a low-lying spin
singlet $\Gamma$.D6.E2 level, and the absence of a low-lying
$\Gamma$.D6.A2 level, thus clarifying the presence of a {\em diamond}
VBS phase in this parameter region.

The spectral features of the VBS phase can be detected across various
system sizes from $N=24$ to $N=48$ (only selected sectors) as shown in
\cref{fig:vbs_spectra_panel}. The lowest excited state on all clusters
we studied belongs to the same $M$ momentum and space group sector,
consistent with the diamond VBS order in the thermodynamic limit. 
Hence, the evidence for a VBS is not only robust in the dimer correlations 
in \cref{fig:vbs_dimercorr_panel} for system
sizes $N=36$ and $N=48$ but also in the energy spectra for all system
sizes we studied.

\paragraph{Spin nematic-plaquette phase ---}

\begin{figure}[t]
  \includegraphics[width=\linewidth]{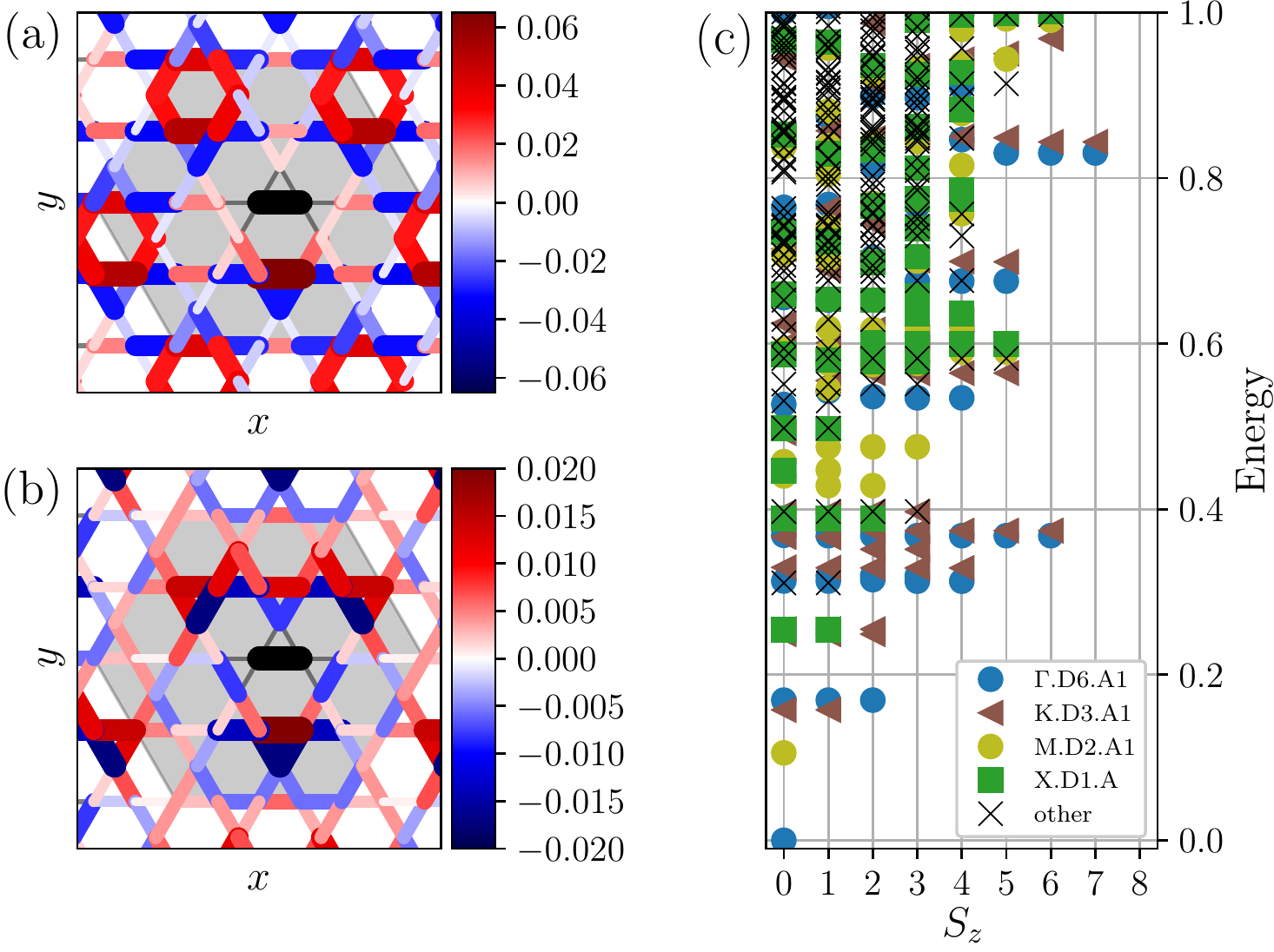}
  \caption{Characterization of a possibly spin nematic phase.  (a): observed
    ground-state dimer-dimer correlations $D_{kl}$ in the nematic
    phase for $J_2/J_1 = -1$, $J_3/J_1 = -2$ as defined in
    \cref{eq:dimercorrs}.  Red (blue) corresponds to positive
    (negative) correlation. The black line is used as the reference
    bond and the grey area shows the $36$-site simulation cell.  (b):
    quadrupolar ground state correlations $Q_{kl}$ as defined in
    \cref{eq:quadrucorrs} showing hexagon-ring sign structure (c)
    Energy spectrum for $J_2 = -1$, $J_3 = -2$ in the nematic phase
    for $S_z < 9$.  The degeneracy at different $S_z$ gives total spin
    quantum number $S$. Odd-$S$ sectors are not present in the
    low-energy tower-of-states, indicating a quadrupolar spin-nematic
    phase.}
  \label{fig:nematic}
\end{figure}

The dimer correlations also exhibit a different peculiar sign
structure in another parameter region, as shown for $J_2=-1$ and
$J_3=-2$ in \cref{fig:nematic}(a). We see characteristic positively
correlated hexagon patterns suggesting a $2 \times 2$ unit cell
superstructure. However we are unaware of a {\em singlet} VBS model
state exhibiting such a correlation pattern. We analogously define an
order parameter for this lattice symmetry breaking pattern,
\begin{equation}
  \label{eq:nematicorderparameter}
  O_{\text{nem}} \equiv \frac{1}{N}\sum_{\langle k,l \rangle} 
  \theta_{kl}^{\text{nem}} D_{kl},  
\end{equation}
where the sign $ \theta_{kl}^{\text{nem}}=\pm 1$ is defined 
in the supplementary material. The region in parameter space where its
signal is strong is shown in \cref{fig:phasediagram}(d).

Since we are unaware of a singlet VBS with this structure, and due to
the vicinity of the ferromagnet, we explore the possibility of a phase
with additional spin-nematic character, for example of quadrupolar
type~\cite{PencLaeuchli2010}. Several examples of frustrated ferromagnets
giving rise to spin nematic phases have been discussed~\cite{Shannon2006,Momoi2006,Hikihara2008,Sudan2009,Iqbal2016}. In
\cref{fig:nematic}(b) we display the quadrupolar bond correlations,

\begin{equation}
  \label{eq:quadrucorrs}
  Q_{kl} \equiv
  \langle ( S^+_k S^+_l) ( S^-_1 S^-_0)\rangle,
\end{equation}
exhibiting sizeable correlations. We notice a peculiar hexagon-ring
sign structure, where the correlations on hexagons surrounding the
middle hexagon are negative, while correlations on the other hexagons
are positive.  In \cref{fig:nematic}(c) we show an energy spectrum
resolved by total $S^z$ and we can see a low-lying $S=2$ level, which
could be due to the quadrupolar character. The lowest singlet excited
state is a M.D2.A1 level, which is in agreement with the reported $2\times 2$
hexagon plaquette superstructure. So we see quite strong evidence for
a novel phase, distinct from the other reported phases, but a detailed
characterization of the phase (e.g. a corroboration of the spin nematic character) 
has to be left for future research.

\paragraph{Discussion and Outlook---}

We have explored the appearance of two unexpected phases along the
classical transition line in the $S=1/2$ kagome Heisenberg antiferromagnet with
competing ferromagnetic further neighbor couplings. The first phase is
a {\em diamond} VBS with a twelve site unit cell. This VBS or variants
thereof have been seen in quantum dimer
models~\cite{Huh2011,Capponi2013,Hao2014,Hwang2015,Ralko2018} and
hinted at by fluctuations or weak correlations in quantum spin models
at the nearest-neighbor point ($J_2=J_3=0$) in
Refs.~\cite{Yan2011,Laeuchli2019}. We have now firmly established this
VBS phase in the extended model~\eqref{eq:hamiltonian}. The location
of this VBS phase in the immediate vicinity of the $\mathbf{q}=0$ magnetic
order, and the apparent second-order nature of the phase transition
between the two phases in exact diagonalization, places this
transition into a contender role for an example of a deconfined
quantum critical transition, with possibly deconfined spin excitations
at the transition~\cite{Senthil2004}.  Recent analytical work on the triangular
lattice~\cite{Jian2018} and the analysis of the matching VBS and N\'eel
monopoles in the Dirac spin liquid~\cite{Song2019} combined with our
numerical results renders this scenario at least plausible. It will also be
important to work out the connection between the VBS phase and the
Dirac spin liquid state, which is currently a prime candidate to describe
the kagome antiferromagnet at small antiferromagnetic $J_2$ 
coupling~\cite{Ran2007,Iqbal2011,Iqbal2013,He2017,Liao2017}, before
entering the $\mathbf{q}=0$ magnetic ordered phase.

This part of the phase diagram is then separated by a likely first
order phase transition from the $\sqrt{3} \times \sqrt{3}$
magnetically ordered phase phase and the lattice symmetry breaking
spin nematic phase close to the ferromagnetic phase. The precise
nature of the latter phase is left for future studies.

\begin{acknowledgments}

  We acknowledge useful discussions with Y.-C.~He, G.~Misguich, and
  C.~Wang.  The Flatiron Institute is a division of the Simons
  Foundation. We thank the Austrian Science Fund FWF for support
  within the project DFG-FOR1807 (I-2868).  This research was
  supported in part by Perimeter Institute for Theoretical
  Physics. Research at Perimeter Institute is supported by the
  Government of Canada through the Department of Innovation, Science,
  and Economic Development, and by the Province of Ontario through the
  Ministry of Research and Innovation.  The computational results
  presented have been achieved in part using the HPC infrastructure
  LEO of the University of Innsbruck. The computational results
  presented have been achieved in part using the Vienna Scientific
  Cluster (VSC). We acknowledge PRACE for granting access to 
  "Joliot Curie" HPC resources at TGCC/CEA under grant number 2019204846.
\end{acknowledgments}

\bibliography{vbskagome.bib}

\end{document}